\documentclass[aps,prl,twocolumn,groupedaddress,showpacs]{revtex4}

\usepackage{graphicx}
\usepackage{amssymb}

\def\he4{$^4$He}
\def\hee3{$^3$He}
\def\t0{$T_0$}

\begin{document}

\title{
Vortex Fluid State below an Onset Temperature $T_0$  of Solid $^4$He }
\author{Andrey Penzev, Yoshinori Yasuta, and Minoru Kubota}
\email{kubota@issp.u-tokyo.ac.jp}

\affiliation{Institute for solid State Physics, University of
Tokyo, Kashiwanoha 5-1-5, Kashiwa, 277-8581, Japan }

\date{\today}

\begin{abstract}

Detailed studies of the AC velocity $V_{ac}$ and $T$ 
dependence of  torsional
 oscillator responses of solid $^4$He
are   reported. A characteristic onset temperature $T_0\sim 0.5$ K
is found, below which a significant  $V_{ac}$  dependent change
occurs  in the energy dissipation for the sample at 32 bar. 
A $V_{ac}$ dependence of the "non-classical rotational inertia"
fraction(NCRIF) also
 appears below  $\sim T_0$.  This value of $T_0$
excludes the possible explanation of supersolid by liquid
superfluidity in grain boundaries or  other liquid related
origins. The $log(V_{ac})$ linear dependence was found in NCRIF.
Furthermore, this linear slope  changes in proportion to  $1/T^2$
for $40< V_{ac}<400~\mu$m/s,
 then crosses over  to $\sim 1/T$ for larger $V_{ac}$.
 We discuss    properties
  of the vortex fluid  proposed  by Anderson above $T_c$ and
  below  $T_0$.
\end{abstract}


\maketitle

Since the first report 
of  "non-classical
rotational inertia" (NCRI)  in solid $^4$He samples by Kim and
Chan\cite{kimchan}, confirmation has come from several  torsional oscillator(TO)
experiments\cite{kimchan,  shirahama, rittnerreppy,
rittnerreppy1,ourqfs}, including by 
the present authors. This finding has been discussed in connection to the NCRI 
of a supersolid as originally proposed by
Leggett\cite{leggett}. A review paper  by Prokof'ev\cite{prok} is
valuable for understanding recent work up to December 2006. An
important conclusion is that the observed phenomena seem to be
more complicated than the original proposal  of a 
BEC of vacancies or other imperfections. Much
excitement has been generated by the recent observation of a
remarkably large NCRI fraction, NCRIF, under appropriate
experimental conditions. Rittner and Reppy\cite{rittnerreppy1}
found the NCRIF increased in
 quench-cooled samples as the distance between closely-spaced,
  concentric walls confining the helium was made smaller.
These authors attribute the increase of NCRIF to increased disorder in the sample. 
 NCRIF greater than 20\% of the total 
mass could be achieved\cite{rittnerreppy1},
 indicating simple mechanisms involving only a small fraction of the solid helium
 are not adequate as explanations of the observed new phase.

According to a recent theoretical proposal by P.W.
Anderson\cite{andersonvortexfluid},  the results previously
attributed to NCRI might be caused by
non-linear-rotational-susceptibility, NLRS, on account of features
shared with non-linear-magnetization seen in some underdoped (UD)
cuprate HTSC\cite{wang} below an onset temperature $T_0$ but above
$T_c$, where the resistivity is non-zero. He discusses the linear
dependence of NLRS on $log(V_{ac})$ as evidence for a vortex
fluid (VF)\cite{andersonvortexfluid}. 
 A fundamental background for the VF state is as follows.  The reported occurrence of NCRI above or near 100 mK  is way too high $T$ for the appearance of BEC of any of the known excitations in solid He from the known concentrations, whereas VF state can appear with the help of vortex excitations in lower dimensional (D) subsystems in the solid He, where quantized vortices have much lower energies and are possibly thermally excited as in 2D Kosterlitz-Thouless(KT) systems. 
 VF state is without 3D macroscopic coherence, and does not support superflow.
More recently
 Kojima's group reports\cite{kojima}  a significant change
 occurring below 40 mK in the TO response
time  when excitation 
 $V_{ac}$ is changed. They also report hysteresis below about this temperature,
 possibly an  indication of a real $T_c$.
 Reppy\cite{reppyMinesota}  claims his group observes similar
hysteresis below a corresponding $T$ although their
"NCRIF" is orders of magnitude larger.  Clark et al.\cite{clark} find
NCRIF appearing at much lower $T$ in either ultra pure \he4 or in 
 \he4 single crystals in comparison to that
seen in samples prepared by the
usual blocked capillary method using the usual commercial grade of
\he4, which typically contains about 0.3 ppm \hee3 impurity. Their
saturation NCRIF has been from 0.03\% to 0.4\%\cite{clark}.

An interesting and significant observation is that the 
NCRIF for a  \hee3 concentration of 0.3 ppm may differ by
more than 3 orders of magnitude among  samples prepared under
different conditions\cite{rittnerreppy1,kojima,clark} while the
characteristic temperatures for the phenomena change by no more
than a factor of 2 to 3. For example, the temperature for the energy dissipation peak $T_p$,
 is below or around 100 mK. The onset
temperature, \t0, below which the NCRI fraction begins to appear
has been reported to be 250 mK to 300 mK\cite{kimchan, shirahama,
rittnerreppy, rittnerreppy1}, except for
\cite{ourqfs,clark,balibar}. This implies some low D
subsystem exists in solid \he4 and the characteristic temperatures
are determined primarily by the subsystem local density while the
number density of the  subsystems determines the overall
NCRIF.  The latter may be 
increased by externally induced disorder
\cite{rittnerreppy1}.  
All these observations seem to imply the conditions for the VF state 
are satisfied and \t0 would imply the appearance of the 
 low D "condensate", as also discussed for UD cuprates\cite{wang}.

We investigated \t0 also 
on account of a claim the observed phenomena
might not be an intrinsic property of the solid\cite{balibar} but
instead could be caused by superfluid liquid at the grain
boundaries.   However, experimentally \t0 
has not been established and it is not known what 
 changes at \t0 because NCRI appears  very gradually. This paper 
 describes experiments on rather stable
   \he4 samples for which NCRIF extrapolated to $T=0$ K is quite small,
    that is  NCRIF(0)$< 0.05$\%. We report our 
   observations and discuss the determination of \t0 and  appearance
   of the VF 
   phase below \t0 and above some $T_c$.

The samples studied were at pressures between 32 and 35.5 bar
 and all showed similar behavior except for the absolute value of the
 dissipation.
All samples remained quite stable as long as we kept them colder
than about 700 mK; with this stability we hoped to study the most
fundamental properties of solid \he4.   Most of the presented data
are for a sample which remained reproducible 
throughout 45 days of experiments, but quite representative of all. 
The measurements were performed on the ISSP fast
rotating cryostat\cite{rotcryostat}.  This provided much more
reliable and reproducible data compared to our previous supersolid
experiments\cite{ourqfs} because this cryostat is far more rigid
while also having much more mass, about 10 metric tons, with 
superior vibration isolation. 
In addition, the ability to
rotate the samples is now available and we plan presentation of
results for DC rotation in future publications. The BeCu TO has a
15 mm long torsion rod with  2.2 mm outside diameter and a 0.8 mm
coaxial hole serving as the filling line. The cylindrical sample
cell made of brass is mounted on a BeCu base integral with the torsion
rod, with threaded fitting and sealed with Wood's alloy.
The interior sample space is 4 mm high and has 10 mm diameter.
Below 4.2 K the resonant frequency of the TO is approximately 1002
Hz with $Q \approx 1.7\cdot 10^6$ as determined from the free
decay time constant.
 The samples were prepared by the blocked capillary method from
  \he4 gas of commercial purity ($\approx0.3$ ppm \hee3) with cooling along
  the melting curve at the rate $\approx 2-5$  mK/min.
No special annealing was attempted but the samples were cooled
slowly, over a period of a few hours, from the melting curve to 1
K.  The final pressure of  solid was estimated from a sharp drop
in TO amplitude at the melting temperature measured during slow
($\approx 0.55$ mK/min) heating after completion of the
measurements\cite{ourqfs}. The change of period caused by the
solidification of the sample  is $\Delta p_{load} \approx 2.4$
$\mu$s.

%
%
%
%
\begin{figure}
\includegraphics[width=0.80\linewidth, bb = 15 17 180 235,clip]
{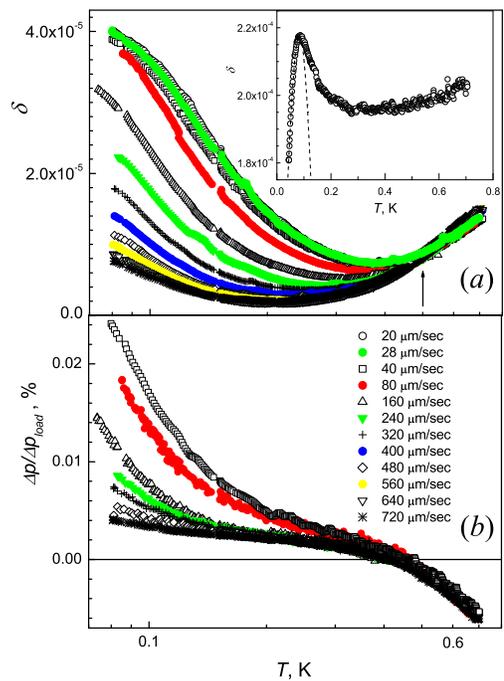} \caption{\label{ncrit}$T$ dependence of energy
dissipation $\delta$({\it a}) and NCRIF= $\Delta p/ \Delta
p_{load}$({\it b})
 at various $V_{ac}$.  The values of $\delta$
are presented without any artificial shift. An arrow indicates $T_0$, across
which $V_{ac}$ dependence changes the sign. 
Some data are omitted for clarity (all the data on $V_{ac}$
dependence are plotted in Fig. 2). The inset in ({\it a})
 indicates a typical energy dissipation peak with  somewhat higher $T_p$.
  The low $T$ part of the peak was fitted  with a Gaussian: dashed
  line. See Fig.2 caption for the determination of zero for NCRIF in ({\it b}).}
\end{figure}
%
%
%
In  order to discuss solid \he4 internal friction  separately from
empty BeCu TO properties we have chosen the quantities associated
with solid \he4 as below, to facilitate 
comparison with results 
from other types of experiments on solid \he4.
Energy dissipation(internal friction) in the solid \he4 sample $\delta$ is evaluated from
 TO measurements taking  similar considerations of the composite TO\cite{composite} where 
  the \he4 sample
  itself is regarded as a part of the composite oscillator and also compared with sound
measurements\cite{tsymbalenko}.
 Using additivity of dissipated energy $\Delta\varepsilon$ and
 the stored energy $\varepsilon$ for the composite TO
per cycle of oscillation, the definition of internal friction
$Q^{-1}=\Delta \varepsilon/2\pi \varepsilon$ gives
\begin{eqnarray}
\Delta \varepsilon_{total}&=&\Delta \varepsilon_{empty}+\Delta
\varepsilon_{solid}, \nonumber \\
\varepsilon_{total}&=&\varepsilon_{empty}+\varepsilon_{solid},
\nonumber \\ \frac{\Delta
\varepsilon_{total}}{\varepsilon_{total}}&=&\frac{\Delta
\varepsilon_{empty}}{\varepsilon_{empty}+\varepsilon_{solid}}+\frac{\Delta
\varepsilon_{solid}}{\varepsilon_{empty}+\varepsilon_{solid}}.
\end{eqnarray}

In our case $\varepsilon_{empty}\gg \varepsilon_{solid}$, because 
the stored energy  $\sim I$ and $I_{empty}\gg I_{solid}$.
Therefore $\delta$ is given as 
\begin{equation}
\delta=\frac{\varepsilon_{empty}}{\varepsilon_{solid}}(Q^{-1}_{total}-Q^{-1}_{empty});
\end{equation} with
\begin{equation}
\frac{\varepsilon_{empty}}{\varepsilon_{solid}}\approx\frac{I_{empty}}{I_{solid}}\approx\frac{p_{empty}}{2\Delta
p_{load}}\approx 210 \mbox{ (for our cell)},\label{cor}
\end{equation}
where $I_{empty}$ and $I_{solid}$ are the moment of inertia of
empty BeCu TO and solid sample respectively.

The upper graph Fig.~\ref{ncrit}({\it a}) shows  $\delta$ in the
solid \he4 sample while the lower graph({\it b})  gives the
relative shift of the period, $\Delta p/\Delta p_{load}$
corresponding to the NCRIF of the solid \he4 as a function of $T$
for various AC cell rim velocities $V_{ac}$.
The inset in Fig.~\ref{ncrit} shows an example of peaks appearing in the data at $T_p$ for samples for approximately the same pressure. The peak is asymmetric as compared with a Gaussian curve fitted to the data on the low $T$ side. All the data in the main graphs are for $T >$$T_p$ for the sample at 32 bar.

It is important to notice both the period and dissipation $\delta$ are changing over
the entire $T$ range for the measurements. Above 0.5K  $\delta$ increases and 
the relative period decreases as $T$ increases. In addition, a
much stronger apparent dependence on $V_{ac}$ begins for $T$ below
0.5 K,  especially for the energy dissipation. The change of the
sign of the $V_{ac}$ dependence allows the assignment of a unique
characteristic temperature $T_0$ = 0.5 K as indicated by an arrow.
%
%
%
At $T>0.5$ K (normal region) the absolute value of $\delta$ can be
compared with available 
data obtained by other
techniques (sound,  elastic deformation). 
The $\delta < 2 \cdot 10^{-5}$ we find is very much smaller than
other available data\cite{paalanen} and the
resonant dislocation vibration mechanism analysis\cite{iwasa}.
 Based on the present size of $\delta$ the most probable mechanism for
  dissipation is thermoelastic internal friction, and not  dislocation
   motion\cite{paalanen} as has been proposed 
for larger excitation experiments.
%
%
%
%
%
%
The original data in Fig.~\ref{ncrit}({\it b})  form a set of parallel
 curves above $\sim 0.45  K$, but for the graph they have been shifted to coincide in this $T$ range. 

A striking difference between the properties seen in
Fig.~\ref{ncrit}({\it a})  in comparison with other superfluid
systems is that the  dissipation peak
is largest for the smallest excitation velocity. 
This behavior is opposite to what is seen for 
a   KT transition\cite{bishopreppy80}, or the superfluid transition in 
 3D He film system\cite{fukuda2005}, or for bulk liquid \he4 in Vycor\cite{ReppyTaylar},
 all systems in which  the 
dissipation increases when the excitation exceeds some critical value including 0.

In the following we consider this unusual  behavior 
as coming from  fluctuations in the VF state, which
is regarded as a kind of superfluid turbulent state. Fluctuations are controlled by external rotation
 \cite{rotturbex} and  characterized by 
a  distribution  
over a certain momentum space\cite{kolmogorov}. The  width of this distribution
is primarily determined by both the longest straight vortex line length, which is of
the order of system size, and the smallest length scale of the vortex tangle or that of
 vortex rings. In other words, what has been claimed to be a critical velocity is actually
 a characteristic velocity of the turbulence. Further analysis 
 made
 this point clearer. 

\begin{figure}[h]
 \includegraphics[width=0.80\linewidth,bb= 15 15 200 260,clip]{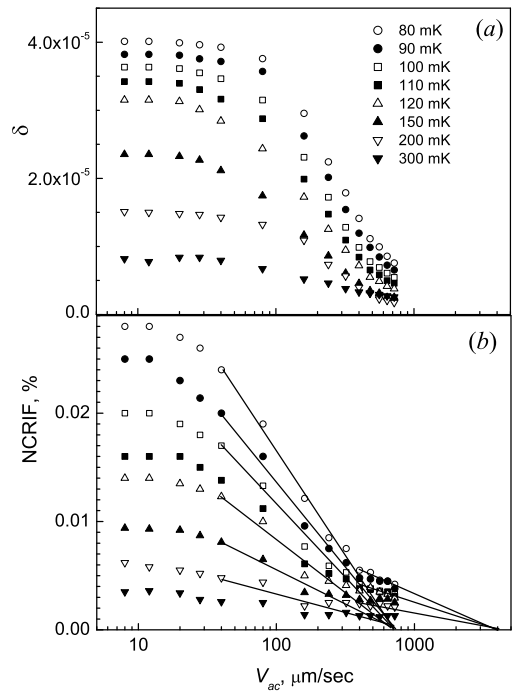}
 \caption{\label{dissipvac}$\delta$({\it a}) and NCRIF({\it b}) as a function of $V_{ac}$
 at $T < 300$ mK.
 The solid lines in ({\it b}) show  the nearly linear dependence on $log(V_{ac})$ for two $V_{ac}$ ranges;
  40$< V_{ac}<$400$\mu$m/s and $V_{ac} > 500$ $\mu$m/s.
   The slope for each range has a unique $T$ dependence given in Fig. 4.   Extrapolated
   lines   are found to converge at  a point for each  $V_{ac}$ range. This point of convergence
    also determines the position of the zero  in Fig. 1({\it b}).}
 \end{figure}



 The  NCRIF and $\delta$ as a function of $V_{ac}$ are
  analyzed at different  $T$'s below 300 mK in Fig. \ref{dissipvac}.
   All the data are taken from the same data set directly from Fig.~\ref{ncrit}.
  If we plot data at higher $T$'s, then we obtain almost horizontal
   displays of data for  each $T$, in the same frame as in (a) for $\delta$ , and the same is
true for NCRIF, but we need to lower the frame bottom to include
higher $T$ data. In Fig.~\ref{dissipvac} NCRIF is constant at low
$V_{ac}$ and starts to decrease above
$\approx 10~ \mu$m/s. 

The most important feature, however, is the linear  dependence on
 $log(V_{ac})$. This dependence was observed previously
in an annular cell \cite{kimchan} and supports the VF
model\cite{andersonvortexfluid}. 
%
%
%
In Fig.~\ref{dissipvac}(b) we observe two velocity regions where
 linear dependence on $log(V_{ac})$ is seen; one from 40 - 400
$\mu$m/sec and the other above 500 $\mu$m/s. We can also
estimate the characteristic     $V_{ac}$  corresponding to 
suppression of the major part of NCRIF as $\sim
750~\mu$m/s. The characteristic velocity for complete suppression
of NCRIF is estimated to be $\sim 4$ mm/s. Moreover, these
characteristic velocities are $T$ independent within our
experimental accuracy. This feature looks
similar to 
$H_{c2}$ of UD cuprate  
\cite{wang}.

\begin{figure}[t]
 \includegraphics[width=1.25\linewidth, bb= 35 25 300 170,clip]
{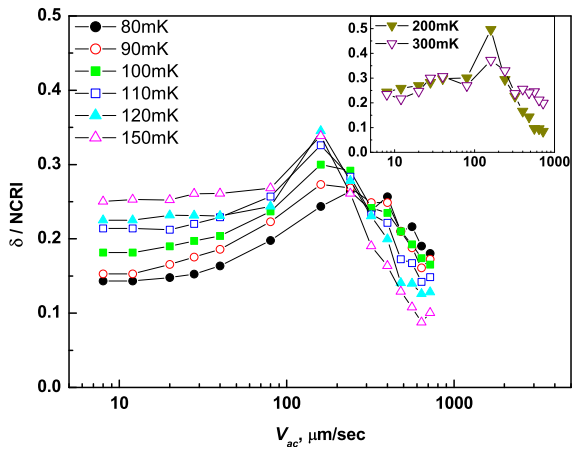} 
 \caption{\label{ratio}Dissipation/NCRIF ratio
  as a function of $V_{ac}$. 
   Apart from zero determination difficulty, we observe a gradual increase at low $V_{ac}$ and
   a wide distribution over more than a decade of $V_{ac}$ with a peak. The peak temperature changes from
    $ \sim 300 \mu$m/s at 80 mK to 
  $\approx 170 \mu$m/s for $T >110$ mK.}
 \end{figure}

\begin{figure}[h]
 \includegraphics[width=0.75
 \linewidth, bb = 0 20 270 210,clip]
 {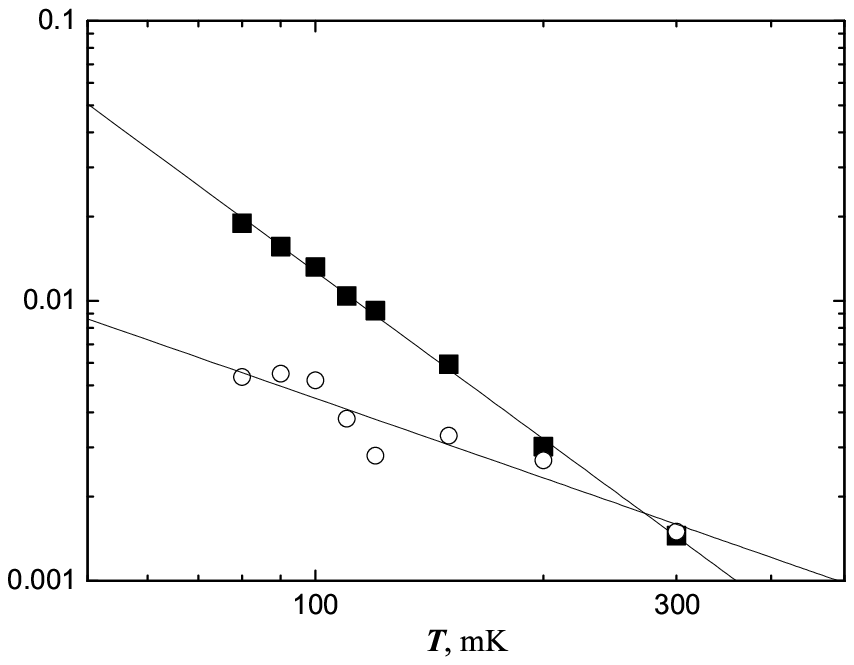}

 \vskip-3cm  \hskip-7cm {\rotatebox[origin=c]{90}{ $\frac{d(NCRIF)}{d(log(V_{ac}))}$}}  \vskip1.5cm

 \caption{\label{slope}T dependence of the slope $d(NCRIF)/d(log(V_{ac}))$.
  Clear $1/T^2$  dependence ($\blacksquare$) is seen for 40 $\mu$m/s$< V_{ac}<400~\mu$m/s and
  a crossover to $\sim 1/T$ for larger $V_{ac}$($\circ$).
  It may correspond to dimensional crossover depending on the length scales of the subsystems.}
 \end{figure}

Another important observation of Fig.~\ref{dissipvac} is the
similarity between ({\it a}) and ({\it b}). In order to study the
energy dissipation
 per superfluid mass, or NCRIF, the ratio $\delta$/NCRIF as a function of $V_{ac}$  for $T < 300$ mK is shown in Fig. \ref{ratio}.
  Despite the uncertainty of
zero for this ratio, we can see
constant level of dissipation at low $V_{ac}$ and then  clear
increase of energy dissipation to some peak value.
The characteristic $V_{ac}$ for the peak position is $\approx 170$
$\mu$m/s for high $T$ and it has a weak temperature dependence;
$\sim 300~ \mu$m/s at 80 mK and changes gradually. We suspect that
this distribution of energy dissipation for all $T< 300$ mK may
correspond to the characteristics of the VF state.
While examining the above evidence we noticed all slopes in the
region 40 $\mu$m/s $< V_{ac}<400~\mu$m/s showed a simple $T^{-2}$
dependence in Fig.~\ref{dissipvac}(b), followed by a crossover to
$\sim 1/T$ dependence as plotted in Fig.~\ref{slope}.
We do not know the origin of these dependences, but it is
interesting to note that it does not include a finite 
temperature shift like Curie-Weiss behavior as for magnetic
susceptibility, but just Curie law like behavior with zero Weiss
temperature. Curie Law behavior is observed for metallic spin
glasses down to the susceptibility peak. This behavior may also
support the idea that we are observing a VF  which freezes at some  $T_c$.
In addition we may have found the involvement excitations of different D origins. 

While preparing 
this paper we discovered 
an interesting study of the mechanical properties in solid \he4 
under shear motion by Beamish's group\cite{shear}. 
We have no concrete idea how solid should behave simultaneously as
superfluid and it will become an interesting question.


In summary, we studied properties of NCRI using TO response from  solid helium samples  at 
 32 bar in wide ranges of $T$ and $V_{ac}$. 
For comparison
with other experimental data, the energy dissipation $\delta$ from solid 
$^4$He itself was evaluated.
 We found $T_0 \sim 0.5$ K from the change of $V_{ac}$
 dependence of  $\delta$. This indicates the appearance of quantized vortices below  $T_0$ and  the
origin of supersolid is not liquid \he4 inside solid. The
suppression of the NCRIF varies nearly  linearly with  $log(V_{ac})$ 
until a crossover to another linear dependence with  $log(V_{ac})$
above $V_{ac} >500~\mu$m/s. The $T$ dependence
changes from $1/T^2$  to $\sim 1/T$. It  looks in 
 support of Anderson's VF picture. 

\begin{acknowledgments}
Authors  acknowledge  T. Igarashi, N. Shimizu and R.M. Mueller's assistance. M.K. is thankful for valuable discussions with P.W. Anderson, D. Huse, and many other
 colleagues in series of workshops organized by Moses Chan and D.~Ceperley, by K. Shirahama,
 as well as by N. Prokof'ev and D. Stamp. Discussions  with M. Kobayashi,
  M. Tsubota and S. Nemirovskii are 
  heartily appreciated. 
  A.P.  thanks JSPS and ISSP for the support.

\end{acknowledgments}

\bibliography{rotationMKshort-1}

\begin{thebibliography}{24}
\expandafter\ifx\csname natexlab\endcsname\relax\def\natexlab#1{#1}\fi
\expandafter\ifx\csname bibnamefont\endcsname\relax
  \def\bibnamefont#1{#1}\fi
\expandafter\ifx\csname bibfnamefont\endcsname\relax
  \def\bibfnamefont#1{#1}\fi
\expandafter\ifx\csname citenamefont\endcsname\relax
  \def\citenamefont#1{#1}\fi
\expandafter\ifx\csname url\endcsname\relax
  \def\url#1{\texttt{#1}}\fi
\expandafter\ifx\csname urlprefix\endcsname\relax\def\urlprefix{URL }\fi
\providecommand{\bibinfo}[2]{#2}
\providecommand{\eprint}[2][]{\url{#2}}

\bibitem[{\citenamefont{Kim and Chan}(2004)}]{kimchan}
\bibinfo{author}{\bibfnamefont{E.}~\bibnamefont{Kim}} \bibnamefont{and}
  \bibinfo{author}{\bibfnamefont{M.~H.~W.} \bibnamefont{Chan}},
  \bibinfo{journal}{Nature} \textbf{\bibinfo{volume}{427}},
  \bibinfo{pages}{225} (\bibinfo{year}{2004}), \bibinfo{note}{~\mbox{Science}
  {\bf 305}, 1941 (2004); Phys. Rev. Lett. {\bf 97}, 115302 (2006)}.

\bibitem[{\citenamefont{Kondo et~al.}(2007)\citenamefont{Kondo, Takada,
  Shibayama, and Shirahama}}]{shirahama}
\bibinfo{author}{\bibfnamefont{M.}~\bibnamefont{Kondo}},
  \bibinfo{author}{\bibfnamefont{S.}~\bibnamefont{Takada}},
  \bibinfo{author}{\bibfnamefont{Y.}~\bibnamefont{Shibayama}},
  \bibnamefont{and}
  \bibinfo{author}{\bibfnamefont{K.}~\bibnamefont{Shirahama}},
  \bibinfo{journal}{J Low Temp Phys} \textbf{\bibinfo{volume}{148}},
  \bibinfo{pages}{695} (\bibinfo{year}{2007}).

\bibitem[{\citenamefont{Rittner and Reppy}(2006)}]{rittnerreppy}
\bibinfo{author}{\bibfnamefont{A.~S.~C.} \bibnamefont{Rittner}}
  \bibnamefont{and} \bibinfo{author}{\bibfnamefont{J.~D.} \bibnamefont{Reppy}},
  \bibinfo{journal}{Phys. Rev. Lett.} \textbf{\bibinfo{volume}{97}},
  \bibinfo{pages}{165301} (\bibinfo{year}{2006}).

\bibitem[{\citenamefont{Rittner and Reppy}(2007)}]{rittnerreppy1}
\bibinfo{author}{\bibfnamefont{A.~S.~C.} \bibnamefont{Rittner}}
  \bibnamefont{and} \bibinfo{author}{\bibfnamefont{J.~D.} \bibnamefont{Reppy}},
  \bibinfo{journal}{Phys. Rev. Lett.} \textbf{\bibinfo{volume}{98}},
  \bibinfo{pages}{175302} (\bibinfo{year}{2007}).

\bibitem[{\citenamefont{Penzev et~al.}(2007)\citenamefont{Penzev, Yasuta, and
  Kubota}}]{ourqfs}
\bibinfo{author}{\bibfnamefont{A.}~\bibnamefont{Penzev}},
  \bibinfo{author}{\bibfnamefont{Y.}~\bibnamefont{Yasuta}}, \bibnamefont{and}
  \bibinfo{author}{\bibfnamefont{M.}~\bibnamefont{Kubota}},
  \bibinfo{journal}{J. Low Temp Phys} \textbf{\bibinfo{volume}{148}},
  \bibinfo{pages}{677} (\bibinfo{year}{2007}).

\bibitem[{\citenamefont{Leggett}(1970)}]{leggett}
\bibinfo{author}{\bibfnamefont{A.~J.} \bibnamefont{Leggett}},
  \bibinfo{journal}{Phys. Rev. Lett.} \textbf{\bibinfo{volume}{25}},
  \bibinfo{pages}{1543} (\bibinfo{year}{1970}).

\bibitem[{\citenamefont{Prokof'ev}(2007)}]{prok}
\bibinfo{author}{\bibfnamefont{N.}~\bibnamefont{Prokof'ev}},
  \bibinfo{journal}{Advances in Physics} \textbf{\bibinfo{volume}{56}},
  \bibinfo{pages}{381} (\bibinfo{year}{2007}).

\bibitem[{\citenamefont{Anderson}(2007)}]{andersonvortexfluid}
\bibinfo{author}{\bibfnamefont{P.~W.} \bibnamefont{Anderson}},
  \bibinfo{journal}{Nature Physics} \textbf{\bibinfo{volume}{3}},
  \bibinfo{pages}{160} (\bibinfo{year}{2007}).

\bibitem[{\citenamefont{\mbox{Wang et al.}}(2006)}]{wang}
\bibinfo{author}{\bibfnamefont{Y.}~\bibnamefont{\mbox{Wang et al.}}},
  \bibinfo{journal}{Phys. Rev. B} \textbf{\bibinfo{volume}{73}},
  \bibinfo{pages}{024510} (\bibinfo{year}{2006}).

\bibitem[{\citenamefont{Aoki et~al.}(2007)\citenamefont{Aoki, Graves, and
  Kojima}}]{kojima}
\bibinfo{author}{\bibfnamefont{Y.}~\bibnamefont{Aoki}},
  \bibinfo{author}{\bibfnamefont{J.}~\bibnamefont{Graves}}, \bibnamefont{and}
  \bibinfo{author}{\bibfnamefont{H.}~\bibnamefont{Kojima}},
  \bibinfo{journal}{Phys. Rev. Lett.} \textbf{\bibinfo{volume}{99}},
  \bibinfo{pages}{015301} (\bibinfo{year}{2007}).

\bibitem[{\citenamefont{Reppy}()}]{reppyMinesota}
\bibinfo{author}{\bibfnamefont{J.~D.} \bibnamefont{Reppy}},
  \bibinfo{note}{private communication.}

\bibitem[{\citenamefont{Clark et~al.}(2007)\citenamefont{Clark, West, and
  Chan}}]{clark}
\bibinfo{author}{\bibfnamefont{A.~C.} \bibnamefont{Clark}},
  \bibinfo{author}{\bibfnamefont{J.~T.} \bibnamefont{West}}, \bibnamefont{and}
  \bibinfo{author}{\bibfnamefont{M.~H.~W.} \bibnamefont{Chan}},
  \bibinfo{journal}{Phys. Rev. Lett.} \textbf{\bibinfo{volume}{99}}
  (\bibinfo{year}{2007}).

\bibitem[{\citenamefont{\mbox{Sasaki et al.}}(2006)}]{balibar}
\bibinfo{author}{\bibfnamefont{S.}~\bibnamefont{\mbox{Sasaki et al.}}},
  \bibinfo{journal}{Science} \textbf{\bibinfo{volume}{313}},
  \bibinfo{pages}{1098} (\bibinfo{year}{2006}).

\bibitem[{\citenamefont{\mbox{Kubota et al.}}(2003)}]{rotcryostat}
\bibinfo{author}{\bibfnamefont{M.}~\bibnamefont{\mbox{Kubota et al.}}},
  \bibinfo{journal}{Physica B} \textbf{\bibinfo{volume}{329-333}},
  \bibinfo{pages}{1577} (\bibinfo{year}{2003}).

\bibitem[{\citenamefont{\mbox{Li et al.}}(2003)}]{composite}
\bibinfo{author}{\bibfnamefont{Z.~S.} \bibnamefont{\mbox{Li et al.}}},
  \bibinfo{journal}{Rev. Sci. Instrum.} \textbf{\bibinfo{volume}{74}},
  \bibinfo{pages}{2477} (\bibinfo{year}{2003}).

\bibitem[{\citenamefont{Tsymbalenko}(1978)}]{tsymbalenko}
\bibinfo{author}{\bibfnamefont{V.~L.} \bibnamefont{Tsymbalenko}},
  \bibinfo{journal}{Sov. Phys. JETP} \textbf{\bibinfo{volume}{47}},
  \bibinfo{pages}{787} (\bibinfo{year}{1978}).

\bibitem[{\citenamefont{Paalanen et~al.}(1981)\citenamefont{Paalanen, Bishop,
  and Dail}}]{paalanen}
\bibinfo{author}{\bibfnamefont{M.}~\bibnamefont{Paalanen}},
  \bibinfo{author}{\bibfnamefont{D.}~\bibnamefont{Bishop}}, \bibnamefont{and}
  \bibinfo{author}{\bibfnamefont{H.}~\bibnamefont{Dail}},
  \bibinfo{journal}{Phys. Rev. Lett.} \textbf{\bibinfo{volume}{46}},
  \bibinfo{pages}{664} (\bibinfo{year}{1981}).

\bibitem[{\citenamefont{Iwasa et~al.}(1979)\citenamefont{Iwasa, Araki, and
  Suzuki}}]{iwasa}
\bibinfo{author}{\bibfnamefont{I.}~\bibnamefont{Iwasa}},
  \bibinfo{author}{\bibfnamefont{K.}~\bibnamefont{Araki}}, \bibnamefont{and}
  \bibinfo{author}{\bibfnamefont{H.}~\bibnamefont{Suzuki}},
  \bibinfo{journal}{J. Phys. Soc. Japan} \textbf{\bibinfo{volume}{46}},
  \bibinfo{pages}{1119} (\bibinfo{year}{1979}).

\bibitem[{\citenamefont{Bishop and Reppy}(1980)}]{bishopreppy80}
\bibinfo{author}{\bibfnamefont{D.}~\bibnamefont{Bishop}} \bibnamefont{and}
  \bibinfo{author}{\bibfnamefont{J.}~\bibnamefont{Reppy}},
  \bibinfo{journal}{Phys. Rev. B} \textbf{\bibinfo{volume}{22}},
  \bibinfo{pages}{5171} (\bibinfo{year}{1980}).

\bibitem[{\citenamefont{\mbox{Fukuda et al.}}(2005)}]{fukuda2005}
\bibinfo{author}{\bibfnamefont{M.}~\bibnamefont{\mbox{Fukuda et al.}}},
  \bibinfo{journal}{Phys. Rev. B} \textbf{\bibinfo{volume}{71}},
  \bibinfo{pages}{212502} (\bibinfo{year}{2005}).

\bibitem[{\citenamefont{Reppy and Tylar}(1991)}]{ReppyTaylar}
\bibinfo{author}{\bibfnamefont{J.~D.} \bibnamefont{Reppy}} \bibnamefont{and}
  \bibinfo{author}{\bibfnamefont{A.}~\bibnamefont{Tylar}}
  (\bibinfo{year}{1991}), \bibinfo{note}{'Excitations in Two-Dimensional and
  Three Dimensional Quantum Fluids', ed. Wyatt and Lauter, Plenum Press, NY,
  pp. 291-300.}

\bibitem[{\citenamefont{Swanson et~al.}(1983)\citenamefont{Swanson, Barenghi,
  and Donnelly}}]{rotturbex}
\bibinfo{author}{\bibfnamefont{C.~E.} \bibnamefont{Swanson}},
  \bibinfo{author}{\bibfnamefont{C.~F.} \bibnamefont{Barenghi}},
  \bibnamefont{and} \bibinfo{author}{\bibfnamefont{R.~J.}
  \bibnamefont{Donnelly}}, \bibinfo{journal}{Phys. Rev. Lett.}
  \textbf{\bibinfo{volume}{50}}, \bibinfo{pages}{190} (\bibinfo{year}{1983}),
  \bibinfo{note}{\mbox{M.} Tsubota, T. Araki, and C. Barenghi, Phys. Rev. Lett.
  {\bf 90}, 205301 (2003)}.

\bibitem[{\citenamefont{Kobayashi and Tsubota}(2005)}]{kolmogorov}
\bibinfo{author}{\bibfnamefont{M.}~\bibnamefont{Kobayashi}} \bibnamefont{and}
  \bibinfo{author}{\bibfnamefont{M.}~\bibnamefont{Tsubota}},
  \bibinfo{journal}{J. Phys. Soc. Jpn.} \textbf{\bibinfo{volume}{74}},
  \bibinfo{pages}{3248} (\bibinfo{year}{2005}).

\bibitem[{\citenamefont{Day and Beamish}()}]{shear}
\bibinfo{author}{\bibfnamefont{J.}~\bibnamefont{Day}} \bibnamefont{and}
  \bibinfo{author}{\bibfnamefont{J.}~\bibnamefont{Beamish}},
  \eprint{arXiv:cond-mat/0709.4666}.

\end{thebibliography}
\end{document}